# An evaluation of DNA-damage response and cell-cycle pathways for breast cancer classification

Atefeh Taherian Fard[1], Sriganesh Srihari[1] and Mark A. Ragan[1][§]

[1]Institute for Molecular Bioscience, The University of Queensland, St. Lucia, Queensland 4072, Australia

[§]Corresponding author: MAR m.ragan@uq.edu.au

# Abstract

## Background

Breast cancer is a highly heterogeneous disease. Accurate subtyping or classification of breast cancer is important for ensuring proper treatment of patients and also for understanding the molecular mechanisms driving the disease. While there have been several gene signatures proposed in the literature to classify breast tumours, these signatures show very low overlaps, considerably different classification performance, and not much relevance to the underlying biology of these tumours. Here we evaluate DNA-damage response (DDR) and cell-cycle pathways, which are critical pathways implicated in a considerable proportion of breast tumours, for their usefulness and ability in breast tumour subtyping. We think that subtyping breast tumours based on these two pathways could lead to vital insights into molecular mechanisms driving these tumours.

## Results

Here, we performed a systematic evaluation of DDR and cell-cycle pathways for subtyping of breast tumours into the five known intrinsic subtypes. We observed that the Homologous Recombination (HR) pathway showed the best performance in subtyping breast tumours, indicating that HR genes are strongly involved in all breast tumours. Comparisons with cell cycle pathway and two standard gene signatures showed that DDR pathways still showed the best performance, thereby supporting the use of known pathways for breast tumour subtyping. Further, the evaluation of these standard gene signatures showed that breast tumour subtyping, prognosis and survival estimation are all closely related. Finally, we constructed an all-inclusive "super-signature" by combining (union of) all genes and performing a stringent feature selection, and found it to be reasonably accurate and robust in classification as well as prognostic value.

## Conclusions

Adopting DDR and cell cycle pathways for breast tumour subtyping achieved robust and accurate breast tumour subtyping, and constructing a super-signature which contains a "good" (feature selected) mix of genes from these molecular pathways as well as clinical aspects (e.g. prognosis and survival estimation) is valuable in clinical practice.

# Background

Tumour cells display a high degree of genomic instability due to the loss of key genes responsible for maintaining the integrity of the genome and suppressing unwarranted cell division [1]. Due to high proliferation rates and genomic instability, tumour cells tend to have different points of genesis and follow different developmental paths, thereby making tumours highly diverse. Such diversity in turn makes it difficult for accurate prognosis and development of effective therapies.

Breast cancer displays highly heterogeneous characteristics [2]. With more than one million reported cases and a mortality rate of 450,000 per year, breast cancer is also one of the most common cancers worldwide, and in 2012 was the most commonly diagnosed cancer among Australian women [3]. This warrants a large-scale study of breast cancer by means of systematic stratification and characterization of subtypes and stages in order to develop effective therapies.

### Breast cancer subtyping

It is critical to classify breast cancer into distinct subtypes for both research purposes as well as in clinical practice. *Clinical subtyping* of breast cancer is usually the first step towards judging the type, dosage and extent of therapy for treating patients. On the other hand, studying and understanding the underlying cellular mechanisms driving breast cancer requires a *molecular subtyping* scheme. While developing a robust subtyping scheme which is relevant for research purposes and is also translatable to the clinic has remained a significant challenge, there have been several proposed schemes, and most of these are based on the expression and mutation of a collection of genes most likely to be involved in the cancer. Here, we focus on classification schemes based on the expression of genes to define molecular subtypes.

Comprehensive gene-expression profiling of breast tumours has revealed at least three major subtypes, namely *luminal (luminal A & B)*, *basal-like or triple-negative*, and the human epidermal growth factor (EGFR) 2 *(HER2)-enriched* subtypes (apart from the less commonly accepted *normal-type*) [2]. Due to morphological and biological differences among breast cancer subtypes, different risk factors, disease prognoses and therapeutic responses are associated with these subtypes [4]. Luminal tumours usually show better response to (hormonal) therapies, and display better prognosis and survival rates. However, triple-negative tumours are highly aggressive and display worse prognosis, and because they do not express any of the three hormone receptors, classical hormonal therapies are not effective in treating these tumours [5].

Gene expression profiling has been extensively used to identify and extract gene signatures for cancer classification, diagnosis and prognosis. Several studies have been undertaken to identify these signatures under different contexts. However, due to lack of a consensus gene signature and thorough understanding of the underlying biological significance of these signatures, these studies have not been fully effective to achieve a robust classification [6].

van't Veer *et al.* [7] developed a 70-gene signature, also known as Amsterdam 70, which is among the most commonly used gene signatures for predicting lymph node negative breast tumours from short intervals to distant metastasis. This signature was identified by comprehensive gene expression profiling of 117 breast cancer tumours, and was further validated on 295 breast cancer patients. The PAM50 gene signature is another commonly used signature for breast cancer prognosis. Developed by Parker *et al.* using microarray and RT-qPCR data generated from 189 breast cancer tumours, a set of 50 genes was selected using the Prediction Analysis Microarray (PAM) algorithm [8] and was shown to have a good predictive capability. On the other hand, Pawitan *et al.* [9] identified a set of 64 genes from 159 breast cancer patients which gave a robust classification between patients with good and poor response to therapies. In another study, Wang *et al.* [10] developed a 76-gene signature by gene expression profiling of 286 lymph node negative breast cancer patients [6]. Several other gene signatures have been developed for the same or similar purposes such as the intrinsic subtype [11-13], recurrence score [14] and two-gene ratio [15] models.

Identifying a standard gene signature for breast cancer classification still remains a challenge. Although the current gene signatures track similar biological characteristics such as prognosis, response to therapy or survival rates, there is very little overlap among them [6]. For instance, the 70 and 76 gene signatures developed by van't Veer *et al.* [7] and Wang *et al.* [10] respectively, are both used for distinguishing metastatic from non-metastatic breast cancer, but share only 3 genes in common [16]. Studies evaluating these signatures have suggested that it is often difficult to understand the underlying biological relevance of these signatures mainly because up to 30% of these genes have an unknown function while the rest of them are associated with unrelated biological pathways [6]. Furthermore, data over-fitting is an inevitable issue while working with high-dimensional data generated from microarray studies, causing gene signatures trained on one dataset to become ineffective in classifying other (independent) datasets [6].

*Contributions of our work*

For the above reasons, we select well-defined *pathways* such as *DNA damage response (DDR) and cell cycle* as a means of differentiating breast cancer subtypes. Constructing gene signatures using known pathways might be a promising approach to overcome these issues in cancer subtyping, particularly understanding the biological significance of the signatures. Moreover, using gene signatures with known biological mechanism can lead to mechanistic insight into the process of disease and eventually testable hypotheses. In fact in a comparative study between different known biological pathways, Liu *et al.* [6] demonstrated that genes involved in cell-cycle pathways could be used as a potential gene signature for breast cancer prognosis. While curated pathways are still incomplete and may miss a few genes, recent significant efforts in KEGG [17], Reactome [18], BioCarta [http://www. biocarta.com] have helped to gather considerable knowledge on key pathways, which when properly adopted could help to bridge this gap between gene signatures, cancer subtyping and biological relevance of signatures and subtypes. Moreover, several studies [19] have suggested that abnormal expression of these genes are associated with DNA damage response in almost all the breast cancer subtypes.

To maintain genome integrity and prevent damage against mutations caused by internal as well as external factors, cells have evolved intricate and robust mechanisms collectively known as DNA Damage Response (DDR). Very broadly, these mechanisms follow three major steps [1]:

(i) sensing DNA damage,

(ii) assembling DNA repair machinery at sites of DNA damage, and

(iii) repair of damaged DNA.

The following are the six core pathways in DDR that we consider in our study:

(i) Base excision repair (BER)

(ii) Nucleotide excision repair (NER)

(iii) Homologous recombination (HR)

(iv) Non-homologous end joining (NHEJ)

(v) Mismatch repair (MMR)

(vi) Fanconi anemia (FA)

# Results

## *Data sources*

The breast cancer gene-expression dataset was obtained from The Cancer Genome Atlas (https://tcga-data.nci.nih.gov/tcga/) containing 547 samples with the following class labels: 98 Basal, 58 HER2, 232 Luminal-A, 129 Luminal-B and 30 Normal-like.

Genes involved in the six different DDR pathways were collated from the public databases KEGG [17] and Reactome [18] and an "in-house" database curated from the literature [20]; genes involved in these pathways are shown in Table 1. Apart from these, we gathered the cell cycle pathway from Liu *et al.* [6]. Next, we gathered three standard gene signatures from the literature, namely Amsterdam-70, PAM-50 and the one from Sotiriou *et al.* [21], shown in Table 2.

## *Relative performance of DDR pathways in breast cancer subtyping*

Figure 1 shows the relative performance of the six DDR pathways in classifying breast cancer samples into the five class labels (Basal, Luminal-A, Luminal-B, Her2+ and Normal-like) using four different clustering algorithms (with number of clusters $k = 5$) measured by adjusted Rand index (ARI) (see Methods). While we can roughly discern from the figure that all methods showed their best performance on HR, it is difficult to accurately gauge the relative performance of the pathways. Therefore, we used a normalization-based ranking scheme [22] to rank the pathways as follows. For each clustering method, we computed the ARI for all pathways and normalized these values against the maximum, as shown in Table 3. We then summed up these normalized values for each pathway across all clustering methods to obtain a total normalized value, which gave the final ranking for all pathways, as shown in Table 4.

Based on Table 4, we could confirm that the double-strand repair HR pathway showed the best performance in classifying breast cancer samples. This meant that HR genes showed the most differences in terms of their expression values across different subtypes. This can be attributed to the change in expression levels of the two key breast cancer susceptibility genes BRCA1 and BRCA2 housed within the HR pathway, between the subtypes, which is likely to impact the expression levels of other genes in the same pathway. Also interestingly, all pathways showed a normalized value of at least 0.50 (relative to HR), thereby indicating that DDR pathways played a considerable role in breast cancer.

Although Table 4 gives a reasonable overall picture of the performance of different DDR pathways, it is difficult to judge whether the lower performance of a pathway (e.g. NHEJ with normalized value 0.569 relative to HR) indicated lower involvement of this pathway in breast cancer *or* it was involved in only a subset (and not all) of the subtypes. Therefore, we evaluated the capability of DDR pathways for *pair wise classification* of breast cancer subtypes to understand the extent of involvement of each of these pathways in different subtypes.

### Relative performance of pathways in pair-wise classification of breast cancer samples

At a time we only considered samples from a pair of subtypes, and evaluated the performance of each pathway in classifying these samples into the two classes. Figure 2 shows the ARI for different clustering methods (number of clusters $k = 2$; see Methods) when used to classify samples in this pair-wise manner.

From the figure we note the performance on Basal *vs* Normal-like and Her2+ *vs* Normal-like strikingly stand out for all pathways. In other words, all pathways are able to differentiate Basal-like from Normal-like and Her2+ from Normal-like with high accuracy. Two observations can be inferred here (although both observations are related to each other): (i) all pathways show considerable differences in the expression levels of their genes between the two subtypes; and (ii) the two tumour subtypes are considerably different from the one another, that is, Basal-like is very different from Normal-like, and Her2+ is very different from Normal-like. This is not entirely surprising because Basal-like is a highly aggressive subtype characterized by worst prognosis and low survival rates, while Normal-like is far less aggressive [2,4,5]. Similarly, Her2 is also aggressive, and therefore different from the less aggressive Normal-like subtype [23]. From the figure, we also see that Basal-like *vs* Her2+

also showed considerable differences across all pathways. Basal-like is predominantly triple negative, that is, does not express the hormones ER/PR/Her2, while the Her2+ type expresses HER2. Her2+ subtype can be treated using hormonal therapy including Trastuzumab and Lapatinib [24], while for triple-negative subtypes hormonal therapies are not effective [5]. Therefore, although both are highly aggressive, these two subtypes are considerably different from one another in terms of molecular characteristics.

Next, Figure 3 shows plots for relative ranking of different pathways in the pair-wise classification of subtypes (using the normalized ranking procedure as before). While all pathways show highest classification capability between Basal-like *vs* Normal-like, and Her2 *vs* Normal-like (as noted before), while we go down the ranks, the pathways show classification capability in separating *different* subtype pairs. For example, the HR and FA pathways can differentiate Basal from Luminal-A better than the remaining pathways. On the other hand, NER can differentiate between Basal and Luminal-B better than the remaining pathways. These observations mean that genes in HR and FA pathways show more differences in their expression levels between Basal and Luminal-A compared to the genes in other pathways. In other words, HR and FA genes are likely to be more responsible for the inherent differences between Basal and Luminal-A subtypes. Similarly, NER genes are more responsible for the inherent differences between Basal and Luminal-B subtypes.

Going back to Figure 1, we see that the ARI of all DDR pathways ranged roughly between 0.20 and 0.40. Therefore, none of the pathways were able to completely differentiate breast cancer samples into all five subtypes with high accuracy (ARI $\geq$ 0.50). This indicated that DDR genes alone are not sufficient to clearly draw the lines between the subtypes and possibly genes from other processes (e.g. cell cycle) need to be included to obtain a more accurate classification. Therefore, we next repeated our experiments by including cell cycle pathway as well as the different gene signatures into our evaluation.

*Relative performance of DDR and cell cycle pathways and gene signatures*

To evaluate the performance of DDR as a whole against other pathways and signatures, we combined all the genes from the six DDR pathways. Figure 4 shows the performance of DDR, cell cycle and the three gene signatures, Amsterdam-70, PAM50 and Sotiriou *et al.* in classifying breast tumours. It is not surprising to see that PAM50 showed the best performance, which is because PAM50 genes were used as part of the procedure to generate the original class labels in TCGA. However, PAM50 does not give 100% accuracy, which can be attributed to the differences between clustering/classification and other post-processing methods adopted by the TCGA consortium, which we do not adopt here.

Figure 4 shows the overall ranking for these pathways and signatures. Interestingly, DDR genes are ranked second (after PAM50) followed by cell cycle and the two gene signatures. This certainly indicates that DDR genes indeed play a crucial role in breast cancer.

Having said that, we note the two gene signatures, Amsterdam-70 and Sotiriou et al., are not primarily designed for subtyping breast tumours into the five intrinsic subtypes, but instead for prognosis and survival analysis, and yet show reasonably good performance ($\geq 0.60$). Further, the Sotiriou *et al.* [21] signature is primarily for classifying breast tumours based on grades. In their work, Sotiriou *et al.* show that a considerable proportion of grade 1 (low grade) breast tumours are ER+, while a considerable proportion of grade 3 (high grade) breast tumours are ER-, and we note that ER+ tumours are predominantly luminal while ER- tumours are predominantly triple-negative or basal-like [23]. These observations mean that breast tumour subtyping, estimation of aggressiveness and prognosis/survival analyses are closely related. In other words, if we can accurately classify breast tumours, we will also be able to considerably predict the aggressiveness of the tumour and also patient survival.

*Constructing a super-signature*

One limitation often raised in the literature regarding gene signatures is the relatively low overlaps among them [6]. When we overlapped the genes from our pathways and signatures, we too observed the same phenomenon (Figure 5). This makes it challenging to obtain an "all-inclusive" signature which is both robust and accurate.

To understand this phenomenon better, we combined the genes from DDR, cell cycle and gene signatures to construct an "all-inclusive" signature containing 381 genes. We then calculated the ARI for this all-inclusive signature using our four clustering methods. While we had expected considerably low ARI due to the inclusion of significant "noise" in this super-signature, to our surprise we saw that the ARI was not very low (around 0.40) compared to the treating the pathways and signatures separately (Figures 1 and 6). Also, all the methods displayed roughly the same ARI, that is, the methods did not show considerable variance as they showed with individual pathways and signatures. Further, performing a stringent feature selection [25] (Table 6) we arrived at set of 17 genes, which we call the "super-signature", which did not decrease the performance of these methods considerably (Figure 6). Interestingly, some of these genes (Table 6) also showed significant prognostic value in terms of patient survival (Figure 7).

These observations instruct developing a super-signature by selecting a "good" (feature selected) mix of genes from relevant molecular pathways (DDR and cell cycle) and clinical aspects (prognosis and survival estimation), which is both robust and accurate in research as well as clinical practice.

## Discussion

Here, we are dealing with a classification problem namely of classifying the samples based on the breast tumor subtype labels. However, we use clustering algorithms for this purpose by semi-supervising them, that is, by setting the number of clusters to be predicted ($k = 2$ or $k = 5$). This leaves room (our future work) to test these algorithms using pathways and signatures for unsupervised classification and thereby possibly identifying novel subtypes.

## Conclusions

Breast cancer displays highly heterogeneous characteristics [2]. Accurate subtyping or classification of breast cancer is therefore crucial in both clinical practice as well as for research purposes. While there have been several signatures proposed in the literature to classify breast tumours, these signatures show very low overlaps, considerably different performance, and not much relevance to the underlying biology of these tumours [6]. On the other hand, we note that DDR and cell cycle pathways are significantly involved in all or most breast tumours, and therefore these pathways are valuable for breast tumour subtyping. Further, being curated based on biological properties, these pathways provide a better understanding of the underlying mechanisms of these tumours.

Here, we performed a systematic evaluation of DDR and cell cycle pathways, and also compared their performance against standard gene signatures. We observed that DDR pathways showed the best performance in classifying breast tumours into the known five intrinsic subtypes, thereby strongly indicating that DDR genes are considerably involved in these tumours. In particular, we noted that the HR pathway plays a key role in all breast tumours. Further, the evaluation of standard gene signatures, which are primarily developed for prognosis and survival estimation, also showed reasonably good performance, further indicating that breast tumour subtyping, prognosis and survival estimation are all closely related. Finally, we attempted to develop a "super-signature" by combining all the genes and performing a stringent feature selection, and to our surprise found it reasonably accurate as well as robust across multiple clustering methods and also significant in terms of prognostic value. This hints at developing such a super-signature which contains a "good" mix of genes from different pathways (DDR and cell cycle) and clinical aspects (prognosis and survival analysis), and which can be both used for molecular subtyping and also in clinical practice.

## Methods

Computational analyses including visualization were coded in a pipeline using open source libraries in Python programming language. Prior to clustering and data analysis, each input file was normalized and rescaled using z-score (mean = 0 and standard deviation = 1).

Clustering methods namely K-means, average linkage, ward clustering [26] and Hopfield network [27] were applied on the datasets. Pycluster library [28] was used for k-means, average linkage and ward methods. However, Hopfield network was built, trained and utilized from scratch in Python. In each of these clustering methods, we pre-fix the number of clusters $k$ to enable classification.

Scikit-learn library [29] was used for computing the adjusted rand index (ARI). To check which gene signatures are giving the best clustering performance, ARI was then normalized and ranked [22]. The flow diagram of the computational analysis is available in the supplementary website.

**And extended version of this work is published as [31].**

## Availability

The datasets and additional results are available at:

https://sites.google.com/site/breastcancersubtyping/

## Competing interests

The authors declare that they have no competing interests.

## Authors' contributions

ATF and SS conceived and designed the experiments. ATF developed the code and performed the experiments. ATF and SS analyzed the data and wrote the paper. MAR supervised the project and reviewed the manuscript.

## Acknowledgements

We thank Dr Lachlan Coin for the feature selection software [25], Chao Liu for the DDR genes, and Professor Kum Kum Khanna for valuable discussions.
*Funding*: Australian NHMRC grant 1028742 to Dr Peter T Simpson and MAR.

## Figures

**Figure 1: Comparison of DDR pathways clustering performance**
Comparison of DDR pathways capabilities in clustering of breast cancer dataset into 5
subtypes

**Figure 2: DDR pathways clustering performance: a pairwise comparison**

a) HR, b) NER, c) BER, d) MMR, e) NHEJ, f) FA

**Figure 3: Overall rank of DDR pathways in terms of pairwise clustering of
breast cancer subtypes**
Based on normalized total Adjusted Rand Index

**Figure 4: Comparison of clustering performance of gene signatures and
pathways**

**Figure 5 – Overlaps between pathways and gene signatures**

The pathways and gene signature showed very low overlaps, as also noted in earlier
studies [6].

**Figure 6 – Performance of a "super-signature" constructed from pathways and
two standard gene signatures**
   (a) Using all genes (381)
   (b) After stringent feature selection (17)

**Figure 7 – Prognostic value for genes in the "super-signature"**

Kaplan-Meier plots [30] drawn using the online tool KM-Plotter
[http://kmplot.com/analysis/] using number of patients $n = 2878$ for (a) MYB, (b)
CKS1B and (c) BUB1

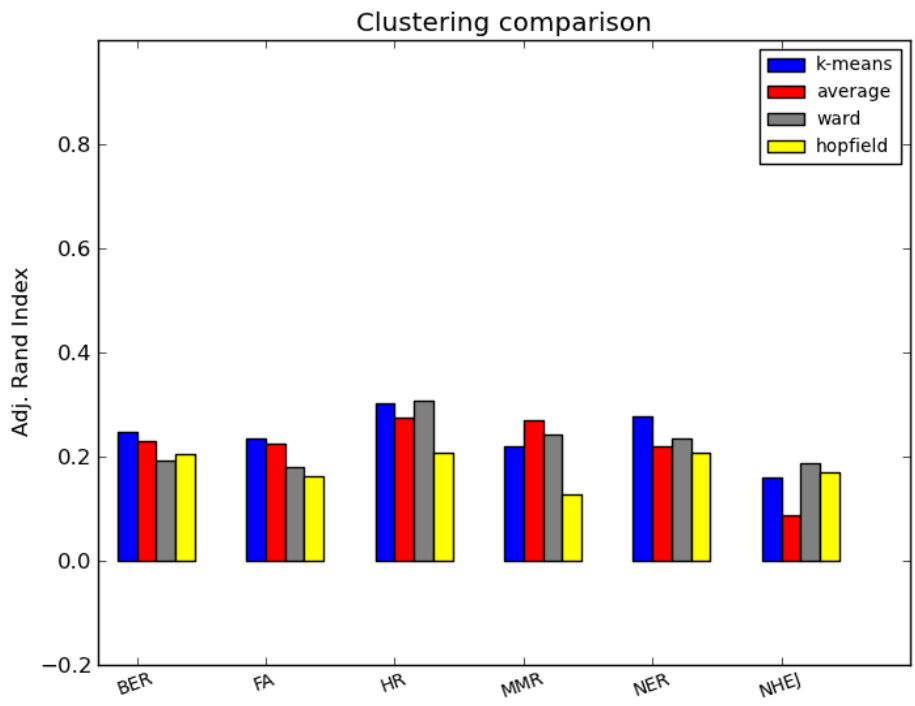

Figure 1

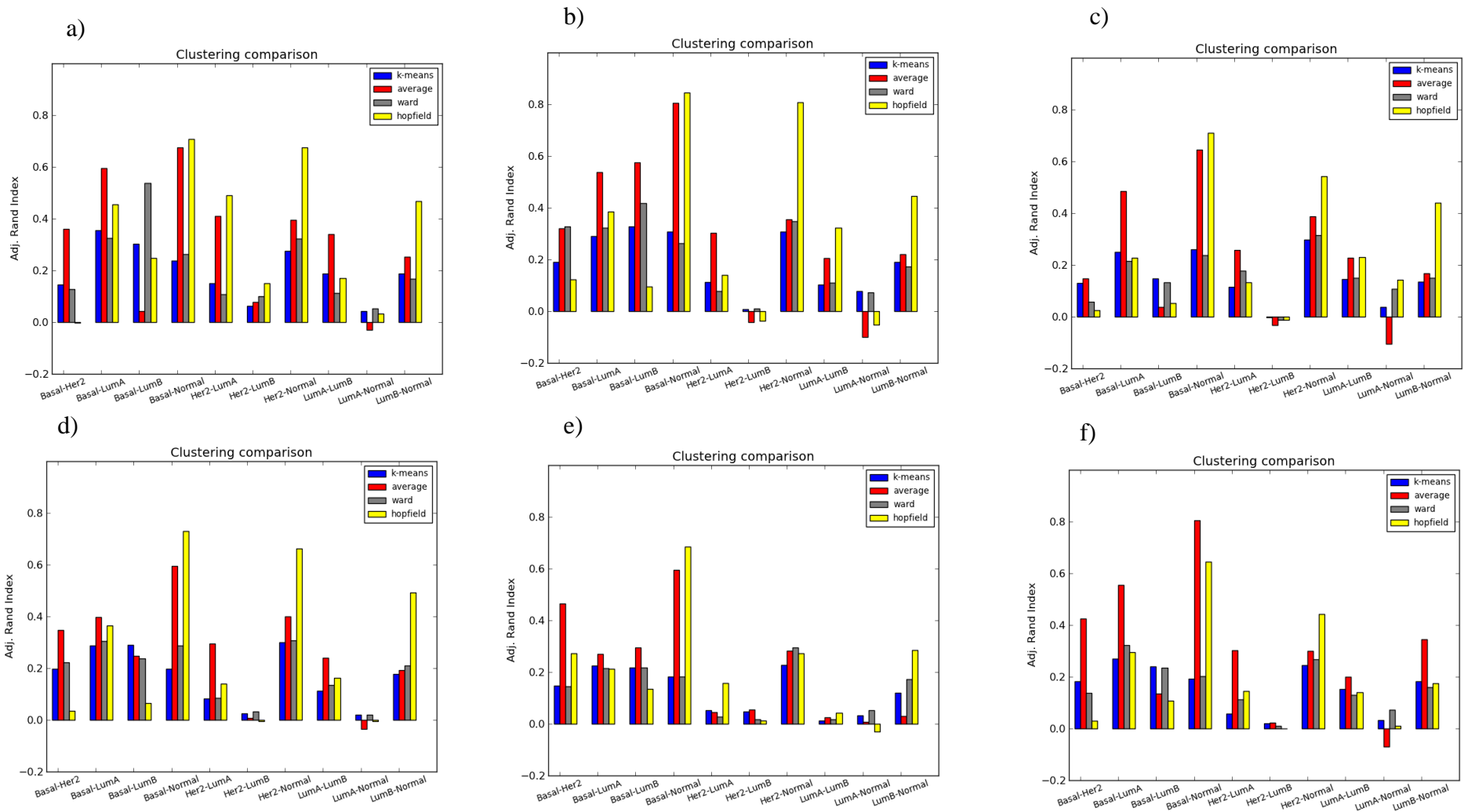

Figure 2

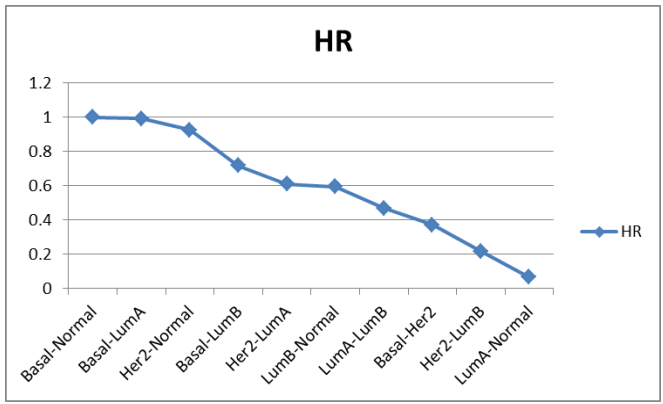

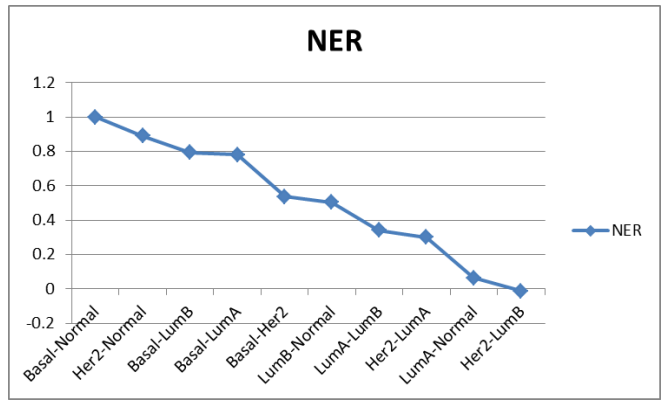

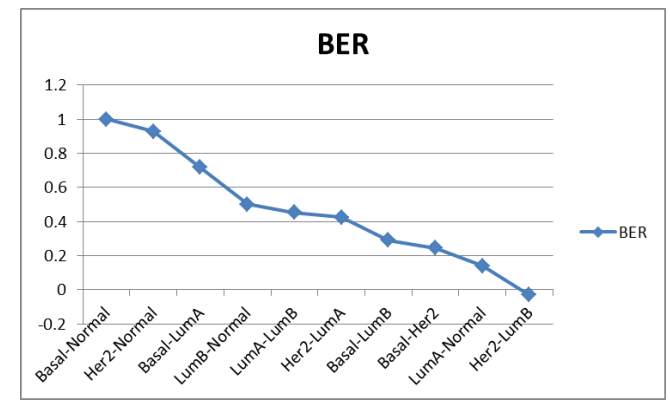

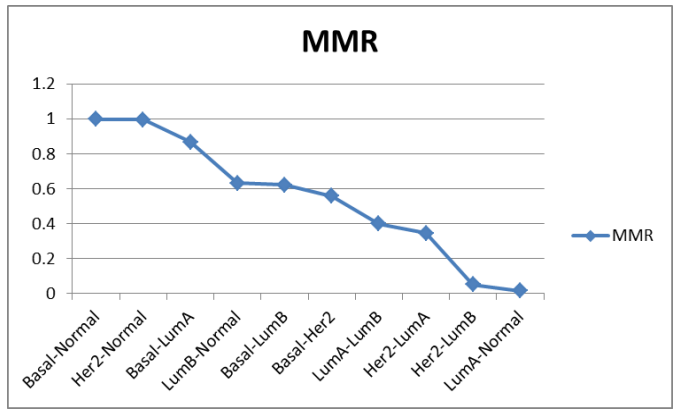

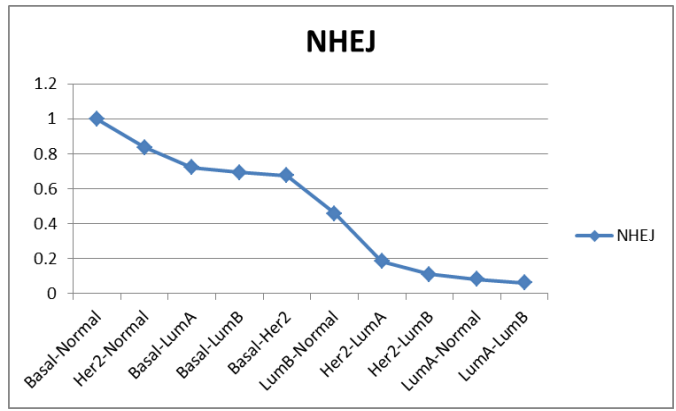

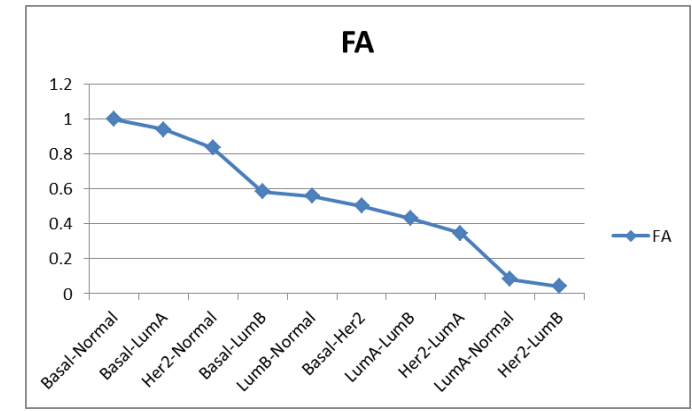

Figure 3

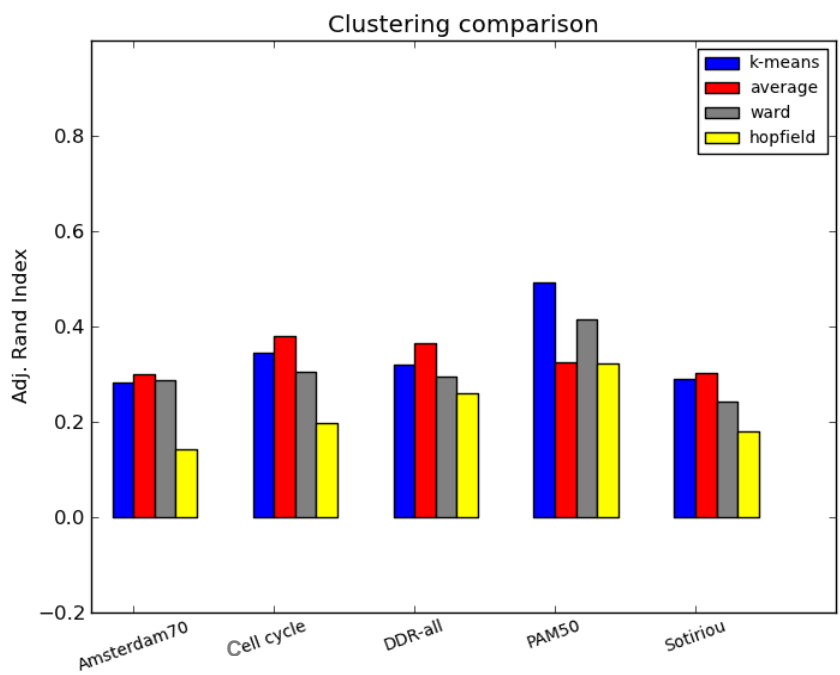

Figure 4

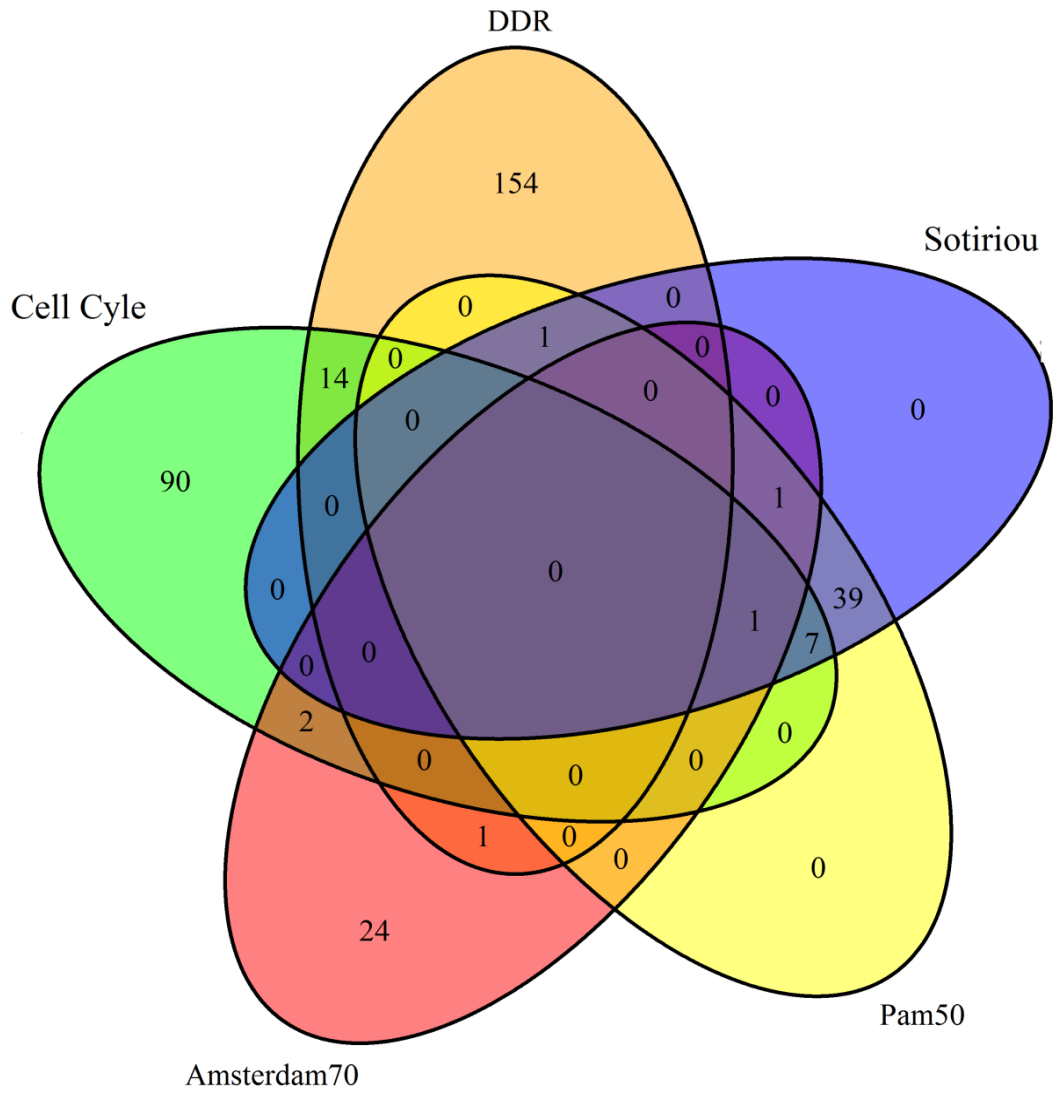

Figure 5

a)

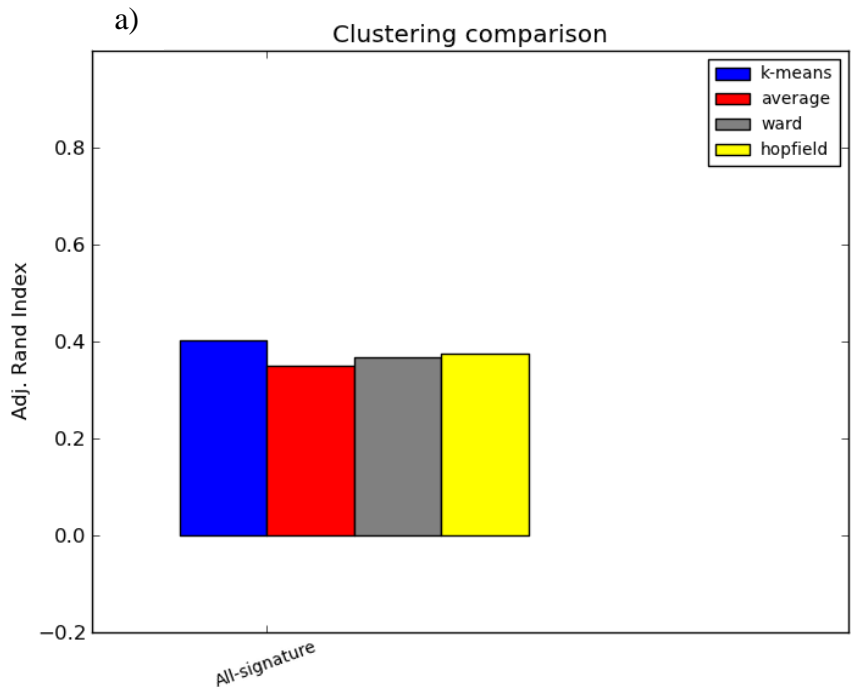

b)

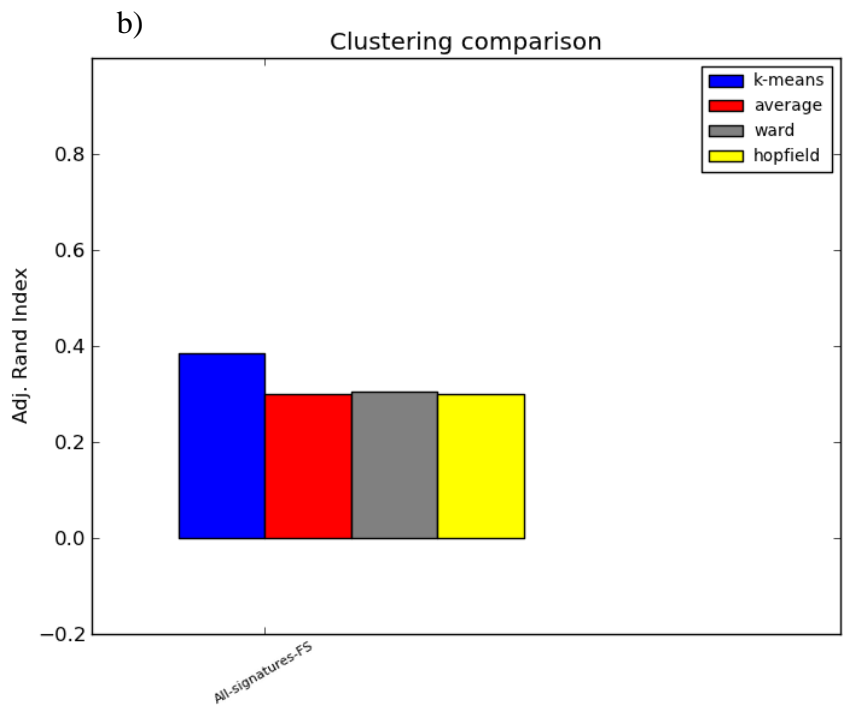

Figure 6

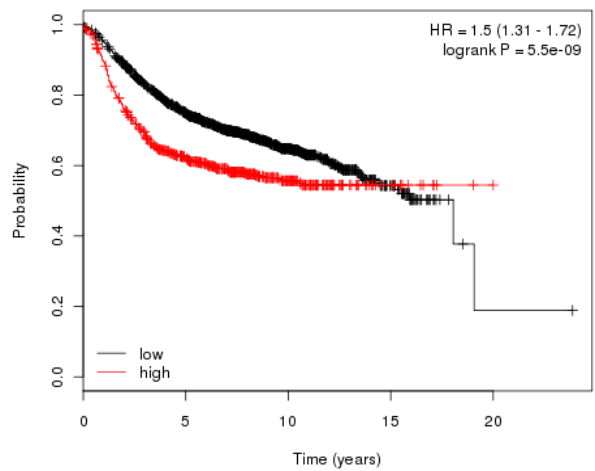

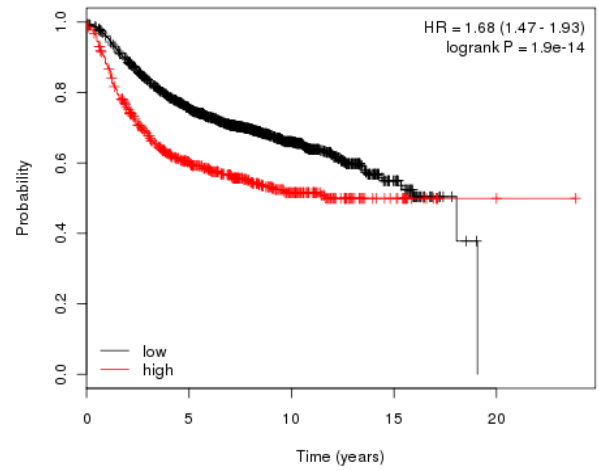

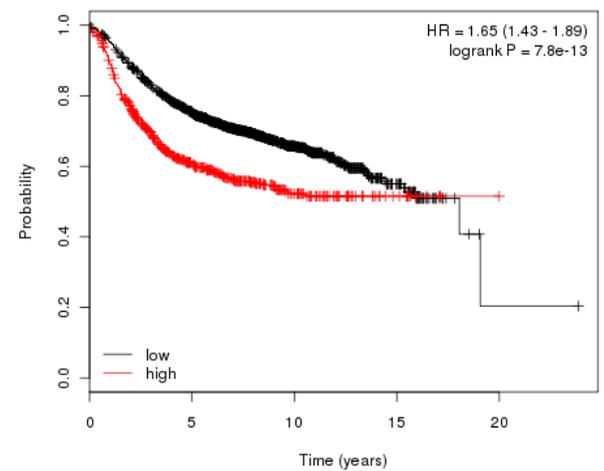

(a)  (b)  (c)

Figure 7

# Tables

## Table 1- List of genes involve in DDR pathways.

| Pathway | Genes |
|---------|-------|
| **BER** | APEX1, FEN1, LIG1, LIG3, MBD4, MPG, MUTYH, NEIL1, NEIL2, NEIL3, NTHL1, OGG1, PCNA, POLB, POLD1, POLD2, POLD3, POLD4, POLE, POLE2,POLE3, POLE4, SMUG1, TDG, UNG, XRCC1, PARP1 |
| **NER** | CCNH, CDK7, CETN2, DDB1, DDB2, EP300, ERCC1, ERCC2, ERCC3, ERCC4, ERCC5, ERCC6, ERCC8, GTF2H1, GTF2H2, GTF2H3, GTF2H4, GTF2H5, HMGN1, LIG1, LIG3, MNAT1, PCNA, POLD1, POLD2, POLD3, POLD4, POLE, POLE2, POLE3, POLE4, POLK, POLR2A, POLR2B, POLR2C, POLR2D, POLR2E, POLR2F, POLR2G, POLR2H, POLR2I, POLR2J, POLR2J2, POLR2J2, POLR2K, POLR2L, POLR2M, RAD23B, RFC1, RFC2, RFC3, RFC4, RFC5, RPA1, RPA2, RPA3, RPA4, TCEA1, TCEA2, TCEA3, XAB2, XPA, XPC |
| **HR** | ATM, ATRX, BARD1, BLM, BRCA1, BRCA2, BRCC3, BRE, BTBD12, C12orf48, C16orf75, C19orf62, CCDC98, CHD4, cPIAS1, CSNK2A1, CSNK2A1P, CSNK2A2, CSNK2B, DNA2L, EME1, ERCC1, ERCC4, EXO1, FLJ40869, H2AFX, HERC2, HTATIP, LIG3, MDC1, MRE11A, MUS81, NBN, OBFC2A, OBFC2B, OTUB1, PALB2, PIAS4, POLD1, POLD2, POLD3, POLD4, POLH, RAD50, RAD51, RAD51AP1, RAD51C, RAD51L1, RAD51L3, RAD52, RAD54B, RBBP8, RBMX, RMI1, RNF168, RNF20, RNF40, RNF8, RTEL1, SHFM1, SLX1A, SLX1B, TOP3A, TOP3B, TP53BP1, TRIP12, UBE2N, UBR5, UIMC1, USP3, XRCC2, XRCC3, PIAS1, RAD51B, RAD51D, RAD54L, RAD54L2, RPA1, RPA2, RPA3, RPA4, SLX4 |
| **MMR** | EXO1, LIG1, MLH1, MSH2, MSH3, MSH6, PCNA, PMS2, POLD1, POLD2, POLD3, POLD4, RFC1, RFC2, RFC3, RFC4, RFC5, RPA1, RPA2, RPA3, RPA4 |
| **NHEJ** | DCLRE1C, LIG3, LIG4, MRE11A, NBN, NHEJ1, PARP1, PRKDC, RAD50, TP53BP1, XRCC1, XRCC4, XRCC5, XRCC6, RBBP8 |
| **FA** | APITD1, BRCA1, BRCA2, BRIP1, C17orf70, C19orf40, EME1, ERCC1, ERCC4, FAN1, FANCA, FANCB, FANCC, FANCD2, FANCE, FANCF, FANCG, FANCI, FANCL, FANCM, MAD2L2,MUS81, PALB2, PCNA, RAD51C, REV1, REV3L, SLX4, STRA13, USP1, WDR48 |

**Table 2- Gene signatures involve in breast cancer classification used in this study.**

| Gene signatures & pathways | Study summary | # of genes | Reference |
|---|---|---|---|
| **Amsterdam70** | Demonstrates a 70 gene expression signature that has a powerful classification capability in 295 breast cancer patients. | 70 | Veer *et al.*[2 in 101] |
| **PAM50** | Developed a 50 gene subtype predictor using microarray and RT-qPCR studies of 189 breast tumour samples indicating good predicting powers in node negative breast cancers. | 50 | Parker J.S. *et al.*[113] |
| **Sotiriou *et al.*** | In a study of 189 invasive breast carcinomas, it was demonstrated that this signatures may improve accuracy of tumour grading and eventually prognosis. | 97 | Sotiriou *et al.*[103] |
| **Cell cycle pathway** | Genes involve in cell cycle were shown to have good predictive capabilities as compare with other gene signatures and pathways, using different breast cancer datasets. | 25 | Liu J. *et al.*[101] |

**Table 3- Relative ranking of DDR pathways clustering capabilities the basis of ARI.**

Relative ranking of DDR pathways clustering capabilities the basis of ARI

| Method | Pathways | RI | Norm |
|---|---|---|---|
| **Kmeans** | **HR** | 0.303 | 1 |
| | **NER** | 0.276 | 0.911 |
| | **BER** | 0.247 | 0.813 |
| | **FA** | 0.234 | 0.773 |
| | **MMR** | 0.221 | 0.727 |
| | **NHEJ** | 0.161 | 0.530 |
| | | | |
| **Average link** | **HR** | 0.276 | 1 |
| | **MMR** | 0.271 | 0.984 |
| | **BER** | 0.229 | 0.830 |
| | **FA** | 0.224 | 0.812 |
| | **NER** | 0.221 | 0.802 |
| | **NHEJ** | 0.087 | 0.315 |
| | | | |
| **Ward** | **HR** | 0.307 | 1 |
| | **MMR** | 0.243 | 0.792 |
| | **NER** | 0.236 | 0.770 |
| | **BER** | 0.193 | 0.628 |
| | **NHEJ** | 0.187 | 0.609 |
| | **FA** | 0.180 | 0.586 |
| | | | |
| **Hopfield** | **HR** | 0.207 | 1 |
| | **NER** | 0.206 | 0.998 |
| | **BER** | 0.204 | 0.985 |
| | **NHEJ** | 0.171 | 0.826 |
| | **FA** | 0.162 | 0.785 |
| | **MMR** | 0.129 | 0.622 |

**Table 4- Gene signatures involve in breast cancer classification used in this study.**

Overall relative ranking of DDR pathways clustering capabilities on the basis of ARI

| Pathway | Total | Norm |
|---------|-------|------|
| HR | 4 | 1 |
| NER | 3.480 | 0.870 |
| BER | 3.256 | 0.814 |
| MMR | 3.125 | 0.781 |
| FA | 2.956 | 0.739 |
| NHEJ | 2.279 | 0.570 |

**Table 5- Gene signatures involve in breast cancer classification used in this study.**

Overall relative ranking of gene signatures clustering capabilities on the basis of ARI

| Signature | Total | Norm |
|-----------|-------|------|
| PAM50 | 3.859 | 1 |
| DDR | 3.128 | 0.810 |
| Cell Cycle | 3.055 | 0.792 |
| Sotiriou | 2.534 | 0.657 |
| Amsterdam70 | 2.501 | 0.648 |

**Table 6- A feature-selected set of genes in the "super-signature"**

List of genes acquired from after feature selection including all gene signatures and pathways in this study:

BUB1, FOXA1, ESR1, WISP1, ERBB2, SLC39A6, CDKN2C, SFRP1, MYBL2, RNF40, KRT5, E2F3, CDC45L, CKS1B, REV1, FGFR4, PGR